\documentclass[useAMS,usenatbib]{mn2e}
\usepackage{times}
\usepackage{natbib}
\usepackage{aas_macros}
\usepackage{amsmath}
\usepackage[pdftex]{graphicx}

\title[]{Radio Halos From Simulations And Hadronic Models I: The Coma cluster}
\author[J. Donnert, K. Dolag, G. Brunetti, R.Cassano, A. Bonafede]{
J. Donnert$^{1}$,
K. Dolag$^{1}$,  
G. Brunetti$^{2}$,
R. Cassano$^{2}$,
A. Bonafede$^{2,3}$\\
$^1$Max Planck Institute for Astrophysics, P.O. Box 1317, D--85741 Garching, Germany\\
$^2$INAF Istituto di Radioastronomia, via P. Gobetti 101, I-40129 Bologna, Italy\\
$^3$Universit\`a di Bologna, Dip. di Astronomia, via Ranzani 1, I-40126
Bologna, Italy}
\begin{document}

\date{Accepted ???. Received ???; in original form ???}

\pagerange{\pageref{firstpage}--\pageref{lastpage}} \pubyear{2008}

\maketitle

\label{firstpage}

\begin{abstract}
We use the results from a constrained, cosmological MHD simulation 
of the Local Universe to predict the radio halo and the gamma-ray flux
from the Coma cluster and compare it to current observations. The simulated
magnetic field within the Coma cluster is the result of turbulent amplification
of the magnetic field during build-up of the cluster. The magnetic seed field
originates from star-burst driven, galactic outflows. The synchrotron emission is 
calculated assuming a hadronic model. We follow four approaches with 
different distributions for the cosmic-ray proton (CRp) population within galaxy 
clusters. The 
radial profile the radio halo can only be reproduced with a radially increasing energy fraction within
the cosmic ray proton population, reaching $>$100\% of the thermal energy 
content at $\approx$ 1Mpc, e.g. the edge of the radio emitting region. Additionally
the spectral steepening of the 
observed radio halo in Coma cannot be reproduced, even when accounting for the 
negative flux from the thermal SZ effect at high frequencies. Therefore the
hadronic models are disfavored from present analysis. 
The emission of $\gamma$-rays expected from our simulated coma is 
still below the current observational  limits (by a factor of $\sim$6) but 
would be detectable in the near future.
\end{abstract}


\section{Introduction}\label{intro}
Galaxy clusters are the largest gravitationally bound objects 
in the Universe.
The thermal gas, which forms the dominant component in the
Intra-Cluster-Medium (ICM), is mixed with 
magnetic fields and relativistic particles, as seen
by radio observations that detected Mpc-sized diffuse radio 
sources in a fraction of X-ray luminous
galaxy clusters in the merging phase 
\citep[e.g.][]{2003ASPC..301..143F,2008SSRv..134...93F}.
A fraction of the energy dissipated during cluster 
mergers may be channelled into the amplification of the
magnetic fields \citep[e.g.][]{2002A&A...387..383D,2006MNRAS.366.1437S} 
and into the acceleration of relativistic, primary
electrons (CRe) and protons (CRp) via shocks and turbulence 
\citep{1998A&A...332..395E,2001APh....15..223B,2007MNRAS.378..245B}.
CRp have long life-times and remain confined within 
clusters for a Hubble time 
\citep[e.g.][and ref. therein]{2007MNRAS.375.1471B}.
Consequently they are expected to be the dominant non-thermal particle
component in the ICM and to generate secondary particles through collisions
with thermal protons.
Primary and secondary particles in the ICM are 
expected to produce a complex emission spectrum from radio to $\gamma$-rays 
\citep[see][ for recent reviews]{2008SSRv..134..191P,2008arXiv0810.0692B,2009arXiv0902.2971C}.

\begin{figure*}
\centering
\includegraphics[width=0.33\textwidth]{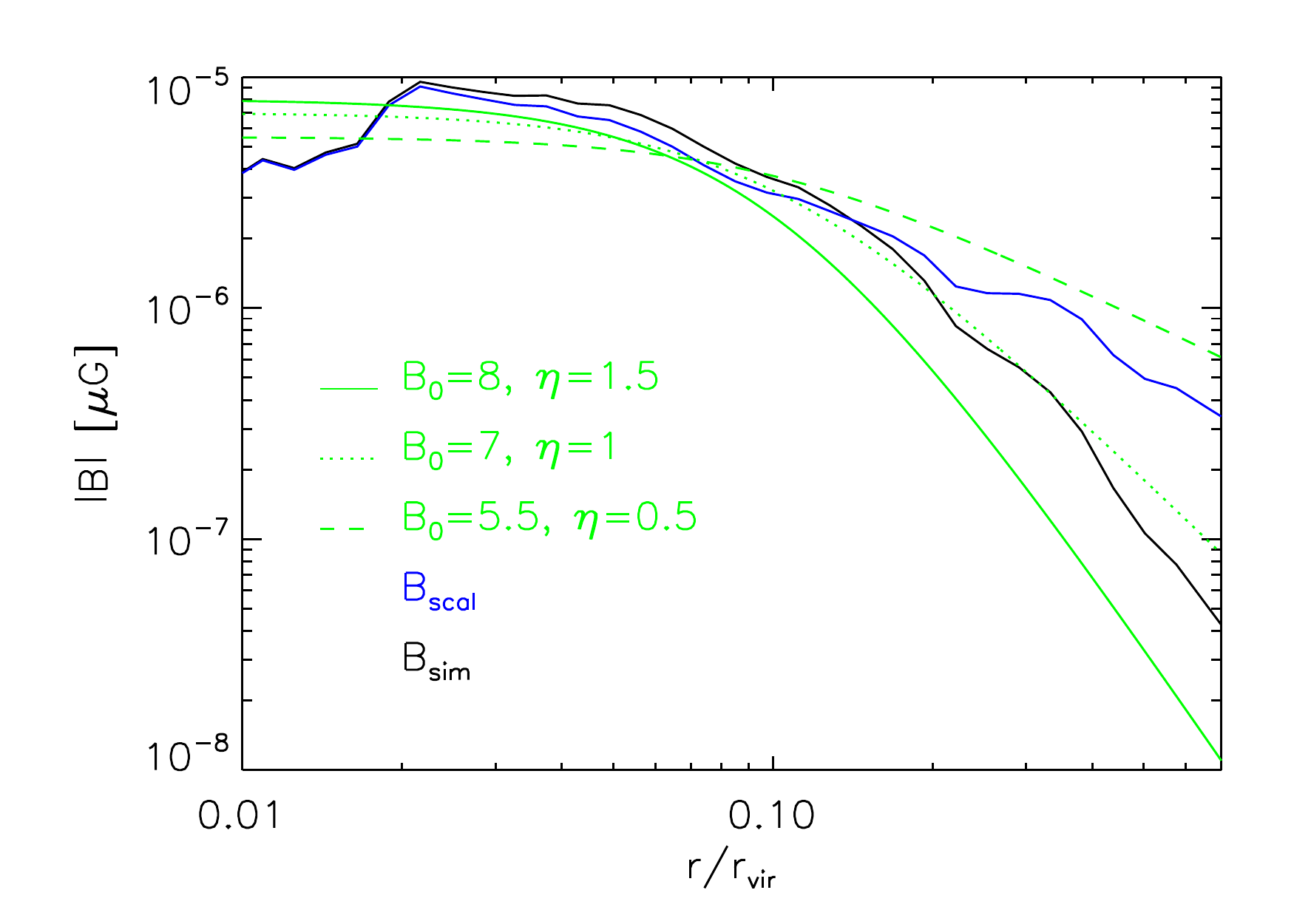}
\includegraphics[width=0.33\textwidth]{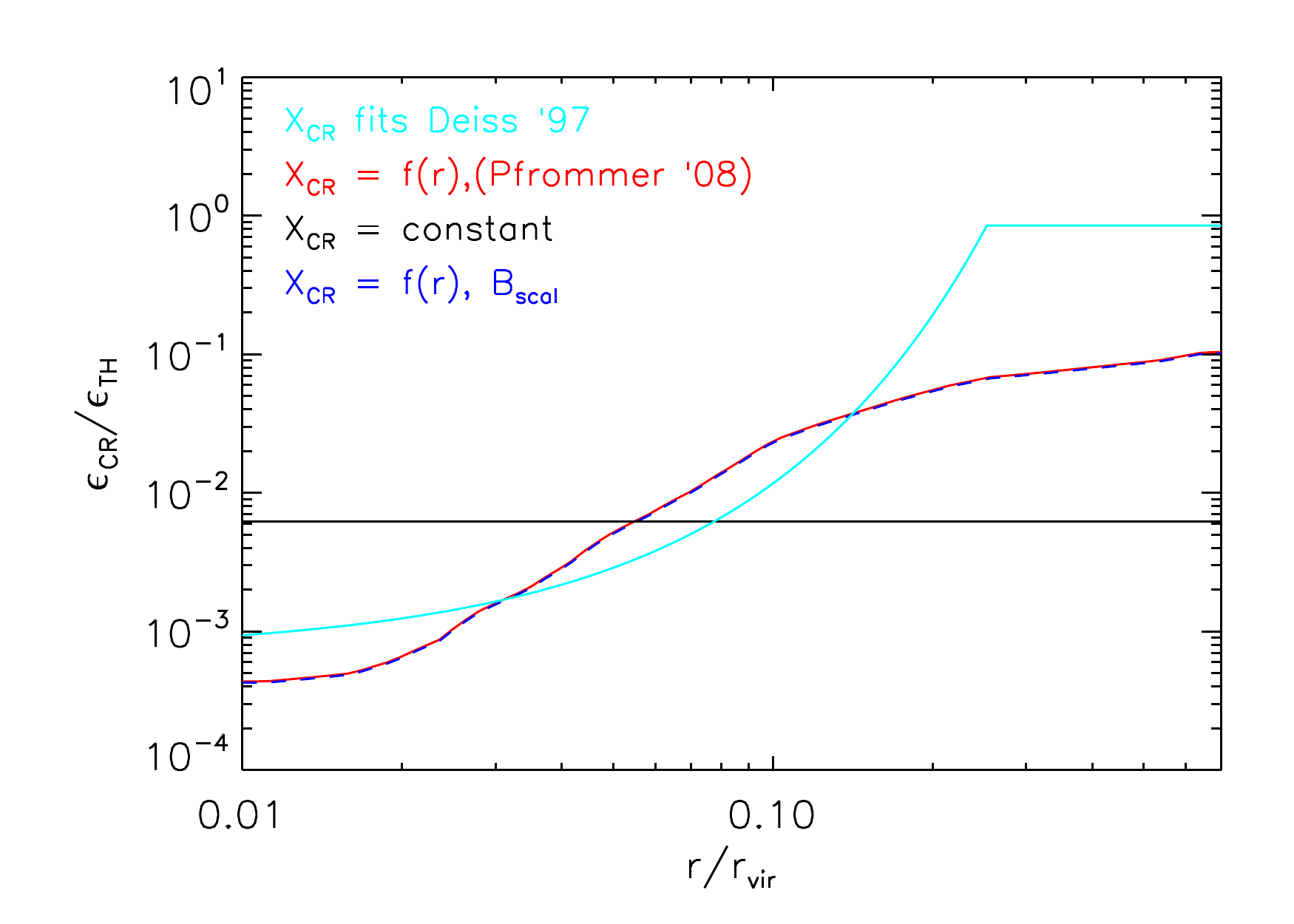}
\includegraphics[width=0.33\textwidth]{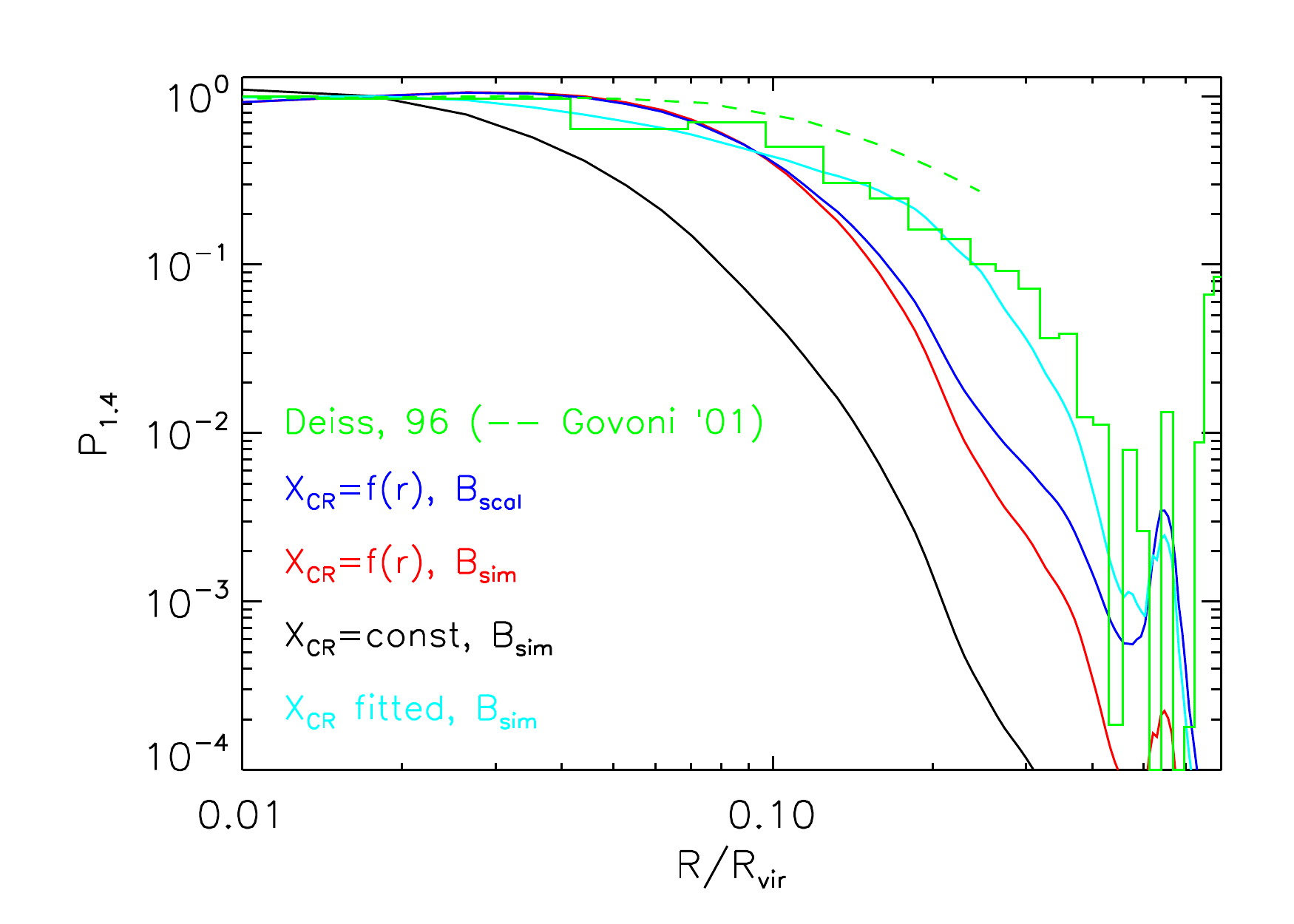}
\caption{The left panel shows the comparison of the radial profile of 
the magnetic field in the simulated Coma cluster (black), 
the scaled version for {\it Model 3} (blue) and the class models 
inferred from the observations. The middle panel shows  
the Energy density fraction of the CRp as function of radius for the 
different models as indicated in the plot. The right pannel shows the 
radial profile for the radio emission resulting of the different models
compared with the observed profile. See text for more details.
}\label{coma_profile}
\end{figure*}

Only upper limits to the $\gamma$-ray emission from clusters have
been obtained so far \citep{2003ApJ...588..155R,2009AA...495...27A},
although the FERMI 
telescope will shortly provide the chance 
to obtain 
the first $\gamma$-ray detections of clusters and to put 
stringent constraints to the energy density of the CRp. 
Future deep observations with high energy Cherenkov arrays are expected to 
provide complementary constraints.
Most importantly, in a few years the Low Frequency Array (LOFAR) and the 
Long Wavelength Array (LWA) will observe clusters at low radio 
frequencies with the potential to discover the bulk of the 
cluster-scale synchrotron emission in the Universe 
\citep[][]{2002A&A...396...83E,2008Natur.455..944B}.

The theoretical picture for $\gamma$-ray emission is very complex and modern 
numerical simulations provide an efficient way to obtain 
detailed models of non-thermal emission from clusters to
compare with present and future observations.
Advances in this respect have been recently obtained by including
aspects of cosmic-ray physics into cosmological Lagrangian 
simulations
\citep{2007MNRAS.378..385P,2008MNRAS.385.1211P}, 
mostly focussing on the acceleration of CRe and CRp at 
shocks and on the production of secondary electrons from such a CRp population.
In this Letter we investigate the non-thermal emission from secondary
particles in a Coma--like cluster extracted from a
cosmological simulation and, for the first time, compare numerical
predictions and observations.

\section{The Simulation}\label{sims}

We use results from one of the constrained, cosmological MHD simulations 
presented in \citet{2008arXiv0808.0919D} from which we select the simulated 
counterpart of the Coma cluster. The initial conditions for a 
constrained realization of the local Universe were the same as used in 
\citet{2002MNRAS.333..739M}.
Briefly, the initial conditions were obtained based on the the
IRAS 1.2-Jy galaxy survey \citep[see][for
more details]{2005JCAP...01..009D}. Its density field was
smoothed on a scale of $7\, \mathrm{Mpc}$, evolved back in time to
$z=50$ using the Zeldovich approximation and used as an 
Gaussian constraint \citep{Hoffman1991} for an otherwise random realization
of a $\Lambda$CDM cosmology ($\Omega_M=0.3$, $\Lambda=0.7$, $h=0.7$). 
The IRAS observations constrain a volume
of $\approx 115 \, \mathrm{Mpc}$ centered on the Milky Way. 
In the evolved density field, many locally observed galaxy clusters 
can be identified by position and mass. The Coma cluster in particular
can be clearly identified by its global properties and gives an 
excellent match to the observed SZ decrement, especially in the non-radiative 
simulations used in this Letter \citep{dolag2005}. 
The similarity in morphology 
of the simulated and observed Coma cluster is coincidental, 
as this structure is far below
the constrains originally imposed by the IRAS galaxy distribution.
The original initial conditions were extended to include gas by splitting dark
matter particles into gas and dark matter, obtaining particle of 
masses $6.9 \times 10^8\; {\rm M}_\odot$ and $4.4 \times 10^9\; {\rm M}_\odot$ 
respectively. The gravitational softening length was set to $10\,\mathrm{kpc}$. 

Our MHD simulation follow the magnetic field through the turbulent amplification
driven by the structure formation process. For the magnetic seed fields
a semi-analytic model for galactic winds was used. Here we used the result 
of the {\it 0.1 Dipole} simulation \citep{2008arXiv0808.0919D}, which 
gives a reasonable match to the observed magnetic field in the Coma cluster.
The left panel of figure \ref{coma_profile} 
compares the magnetic field profile predicted for the Coma cluster 
from our simulations with models that best reproduce the Rotation Measure observed
in five extended sources within the Coma cluster
\citep[preliminary results from ][Bonafede et al. in prep]{2009Bonafede} with a 
magnetic field radial profile : 
$\left<B(r)\right>=B_{0}\left[\frac{n_\mathrm{gas}(r)}{n_0}\right]^{\eta}$ and $\eta$ in the 
range [0.5 -1.5].
Our simulations are in good agreement with a more centrally peaked profile 
(e.g. $\eta \approx 1.0$). This profile gives an 
average magnetic field over the central Mpc$^3$ of $\sim$ 1.9$\mu$G, which is 
consistent with the equipartition estimate derived from the 
radio halo emission \citep{2003A&A...397...53T}. 

\section{Secondary Electrons in galaxy clusters}\label{CRpop}

The origin of Mpc-scale radio emission in galaxy clusters
is still not fully understood.
Extended and fairly regular diffuse synchrotron emission may be produced
by secondary electrons injected during proton-proton collisions, since the
parent relativistic protons can diffuse on large scales (e.g. hadronic or
seconday models; \citet{1980ApJ...239L..93D,1999APh....12..169B}). 
Alternative models assumerelativistic electrons re-accelerated in-situ by MHD turbulence
generated in the ICM during cluster-cluster mergers (e.g. re-acceleration
models; \citet{2001MNRAS.320..365B,2001ApJ...557..560P}).

\begin{figure*}
\center \includegraphics[width=0.24\textwidth]{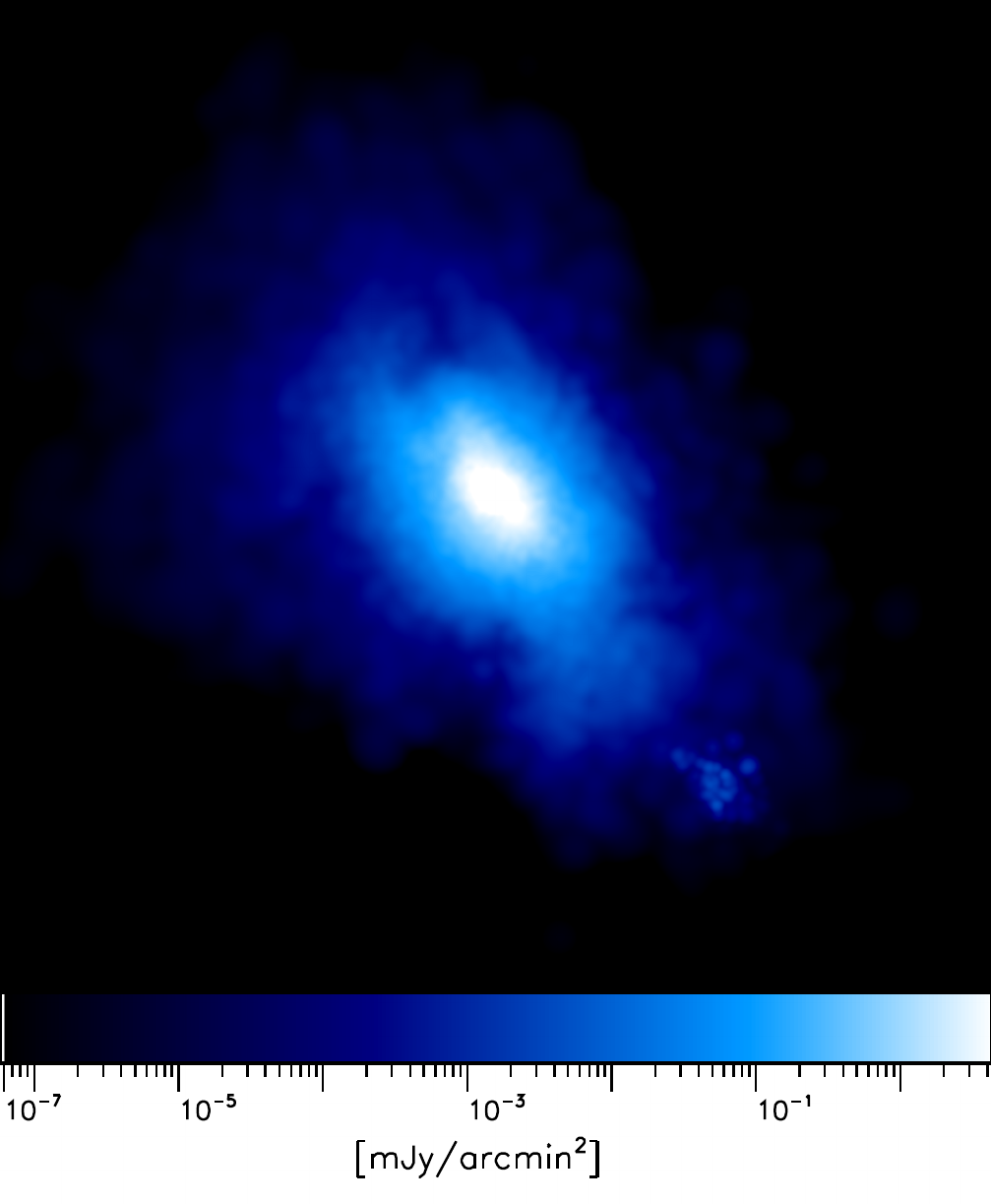}
\includegraphics[width=0.24\textwidth]{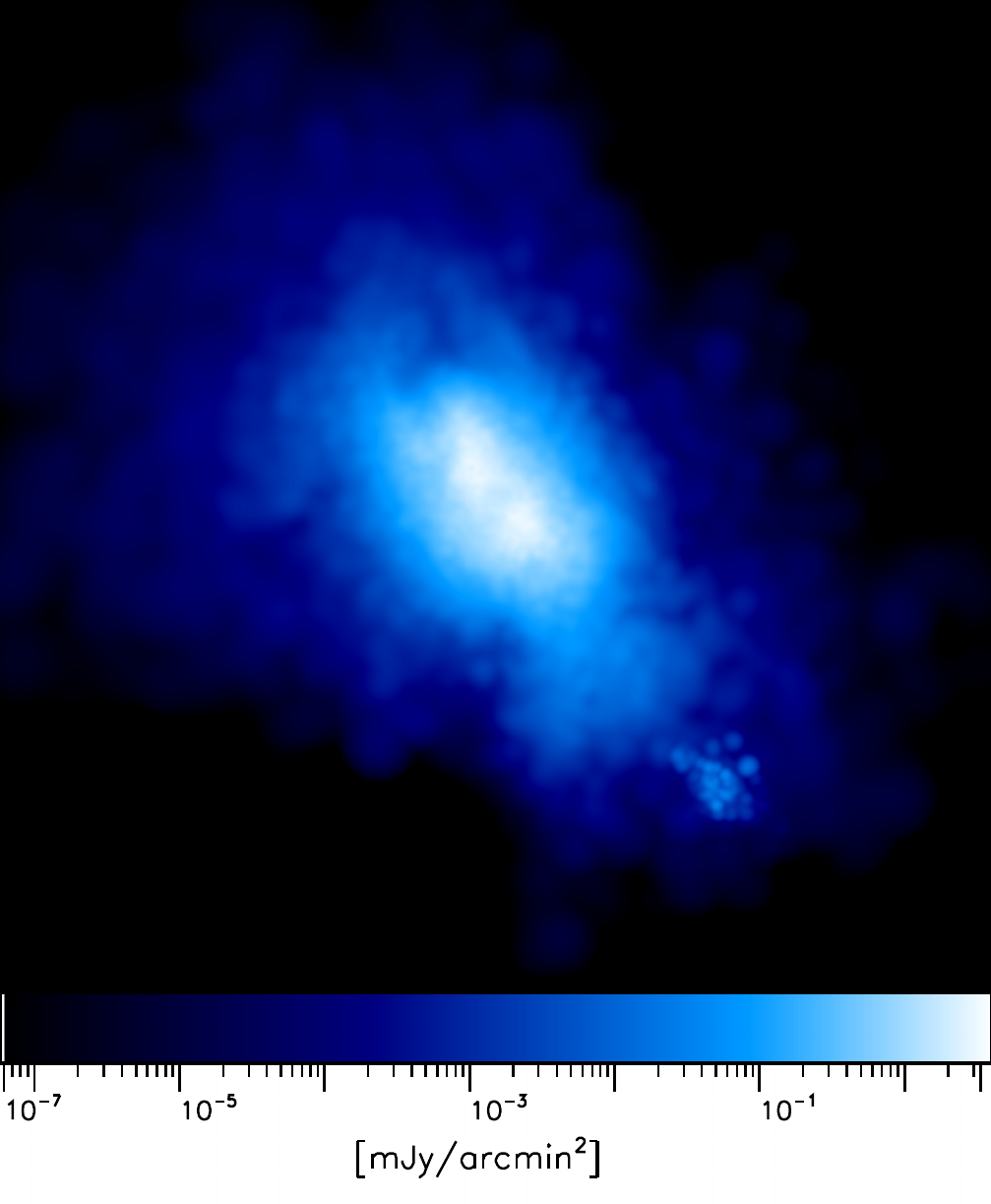}
\includegraphics[width=0.24\textwidth]{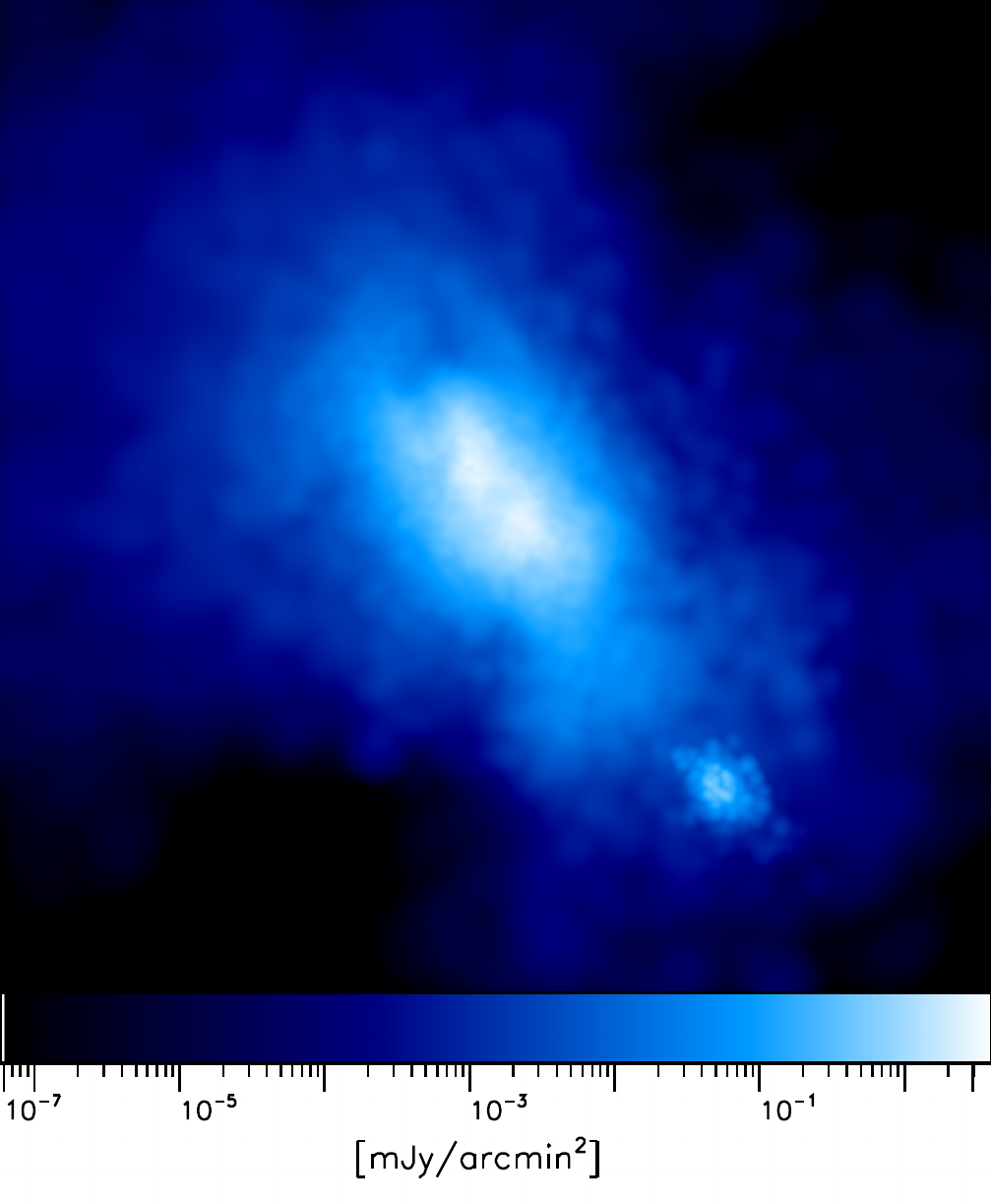} 
\includegraphics[width=0.24\textwidth]{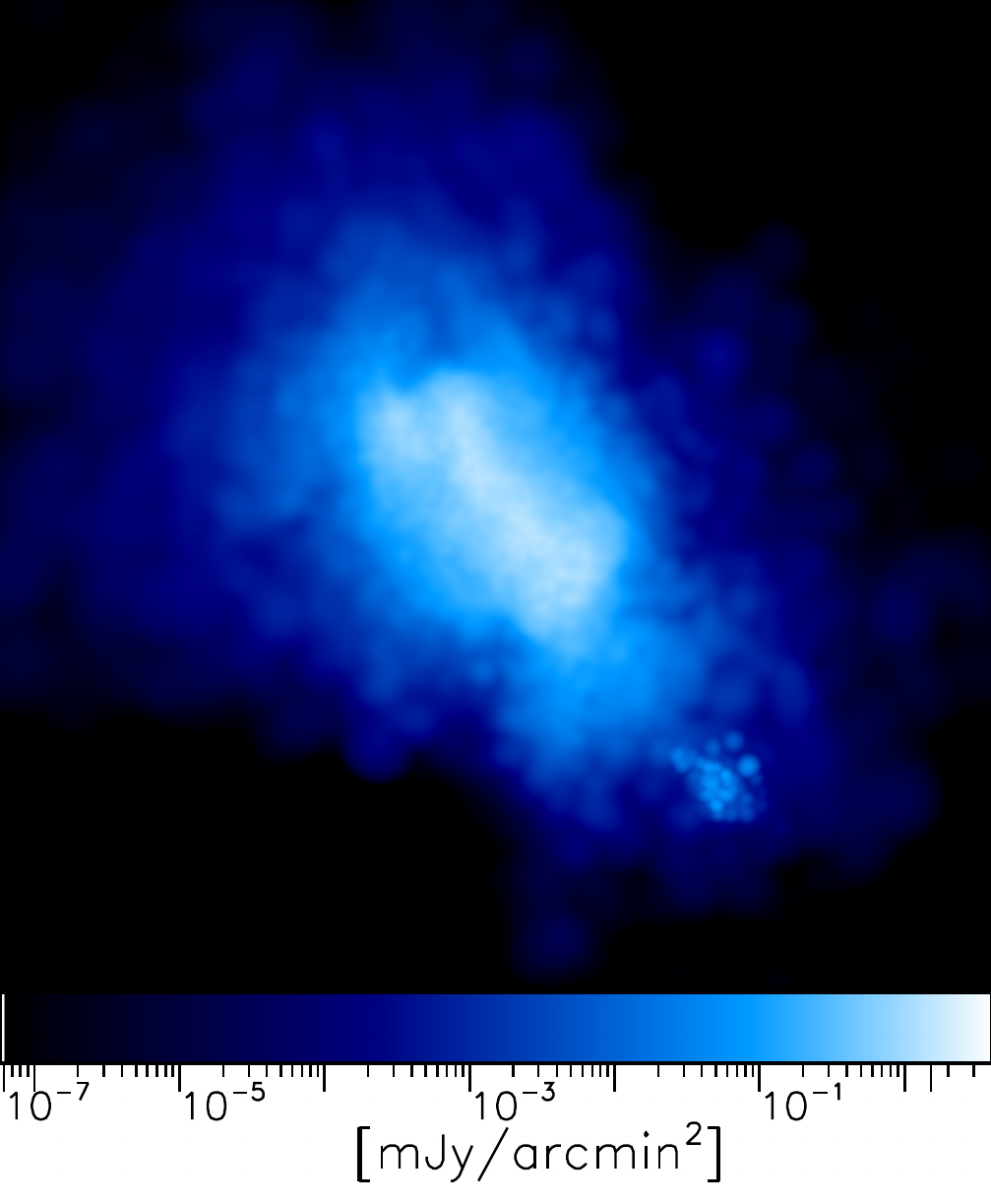} \\

\includegraphics[width=0.24\textwidth]{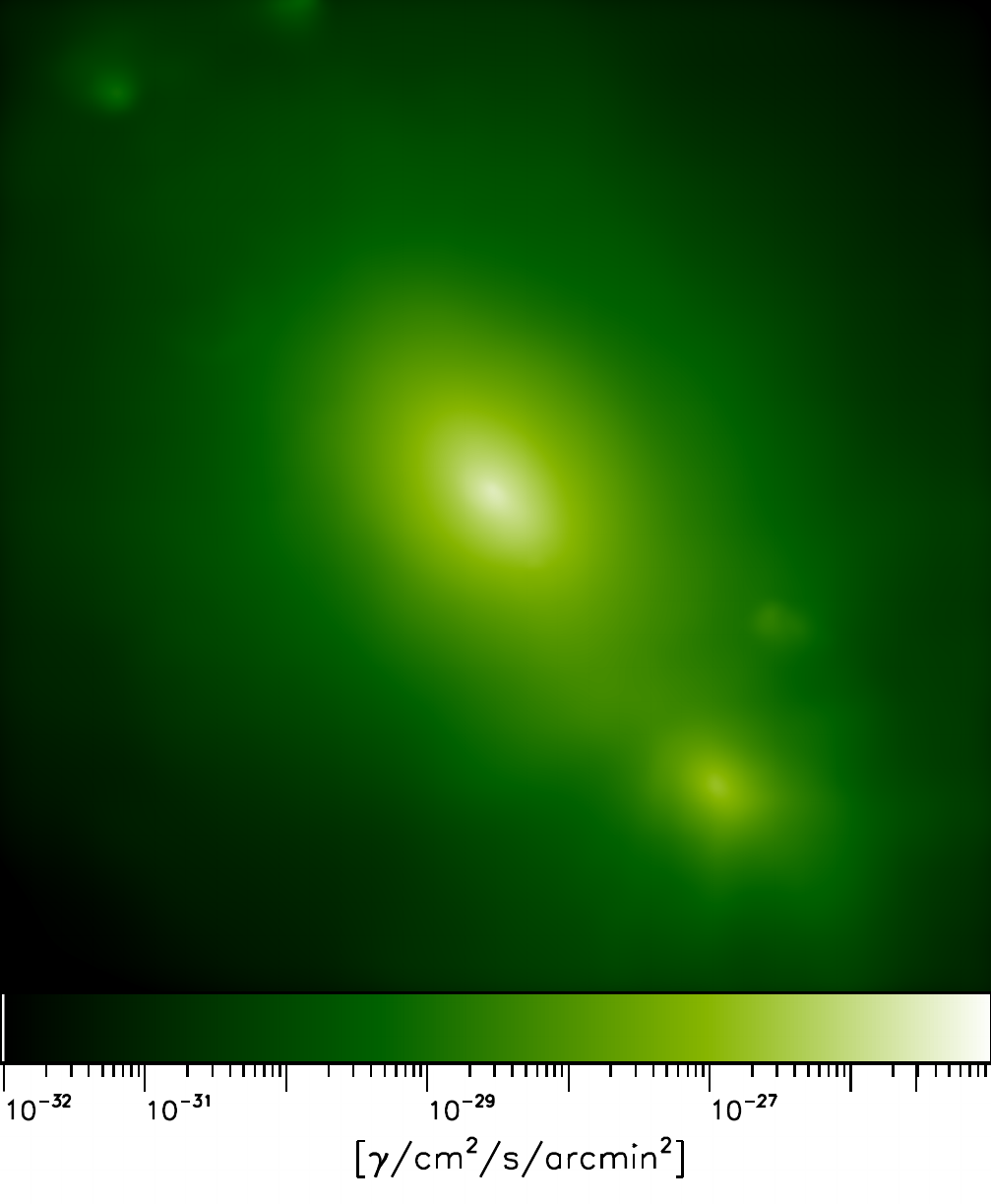}
\includegraphics[width=0.24\textwidth]{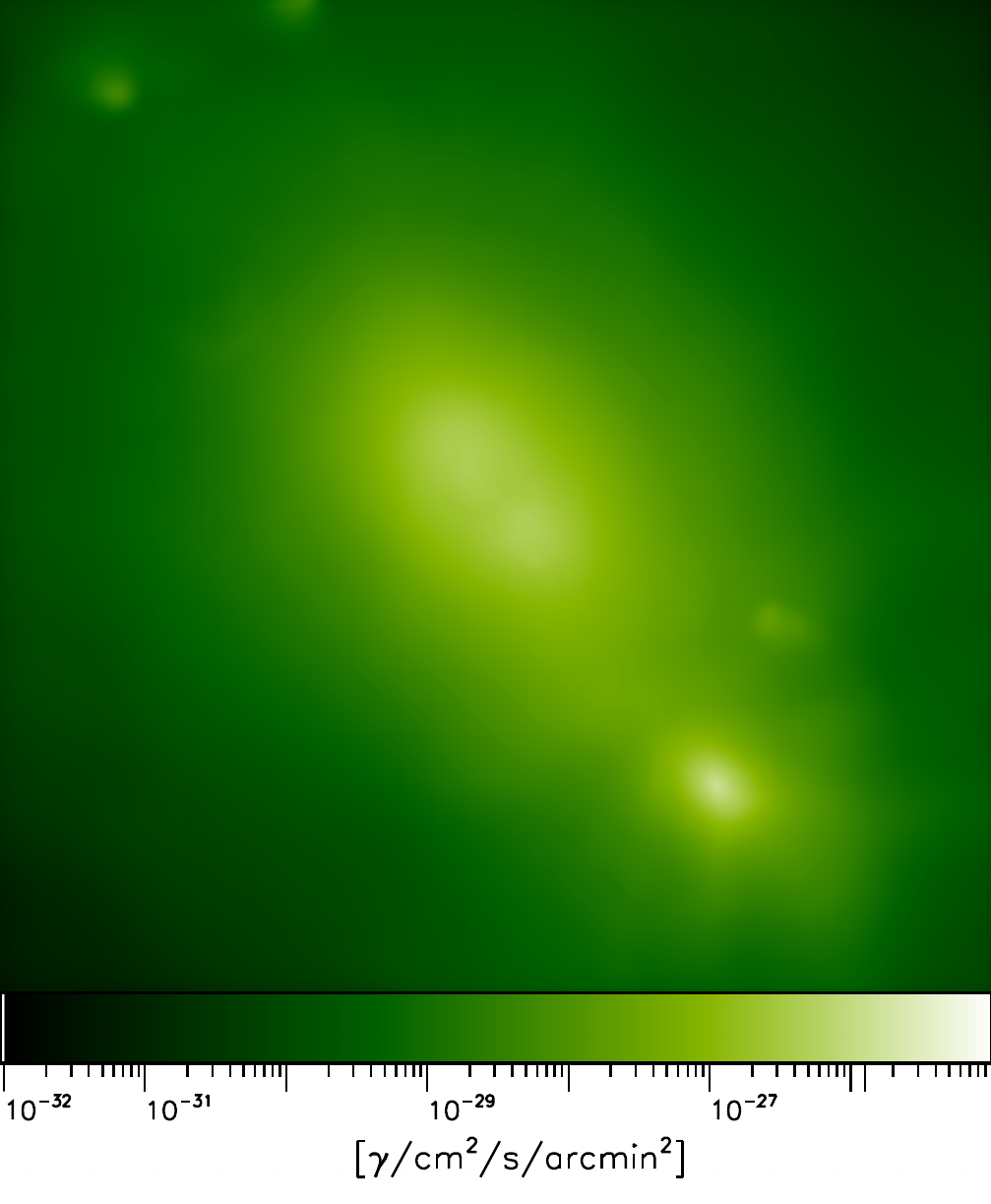}
\includegraphics[width=0.24\textwidth]{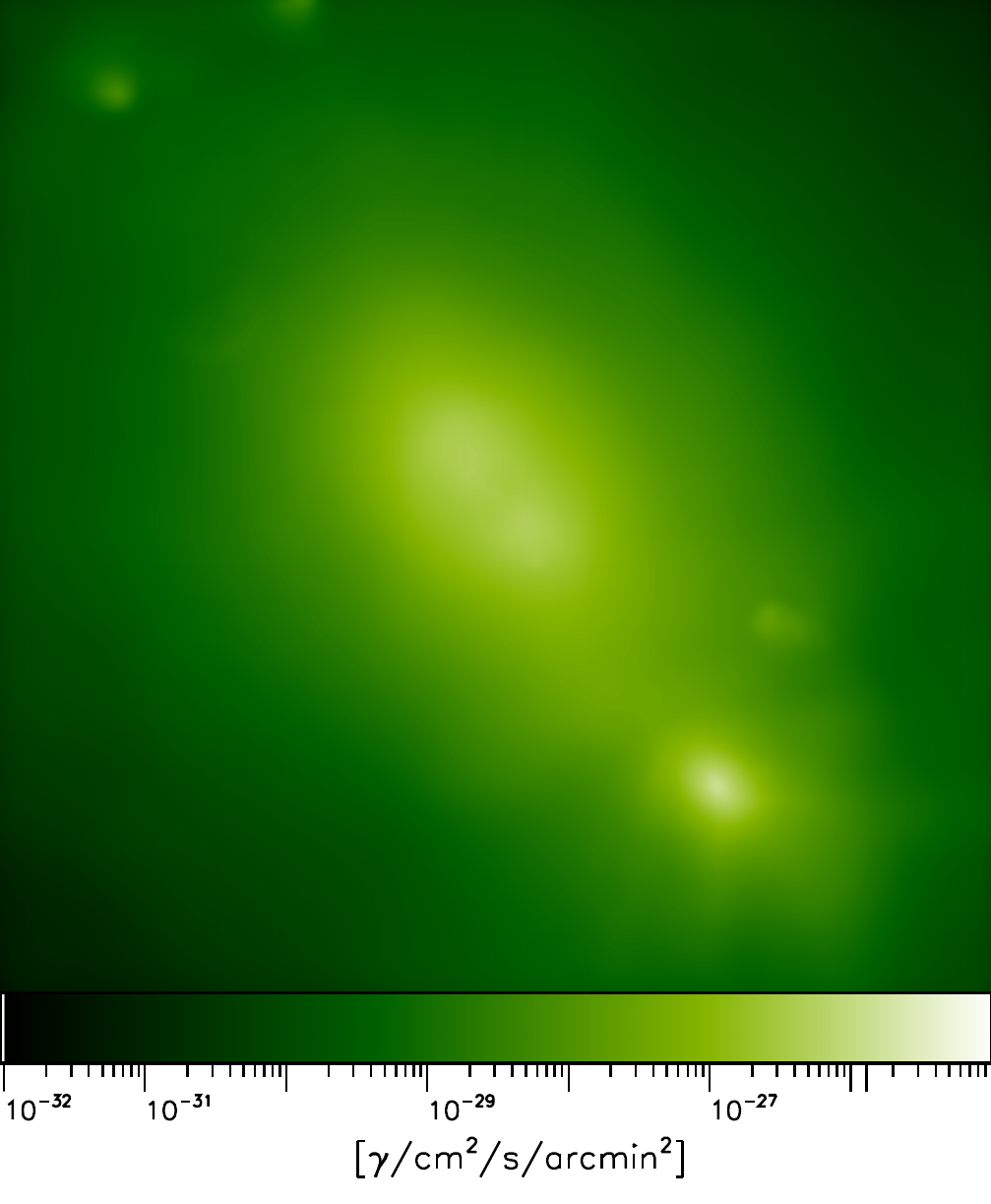}
\includegraphics[width=0.24\textwidth]{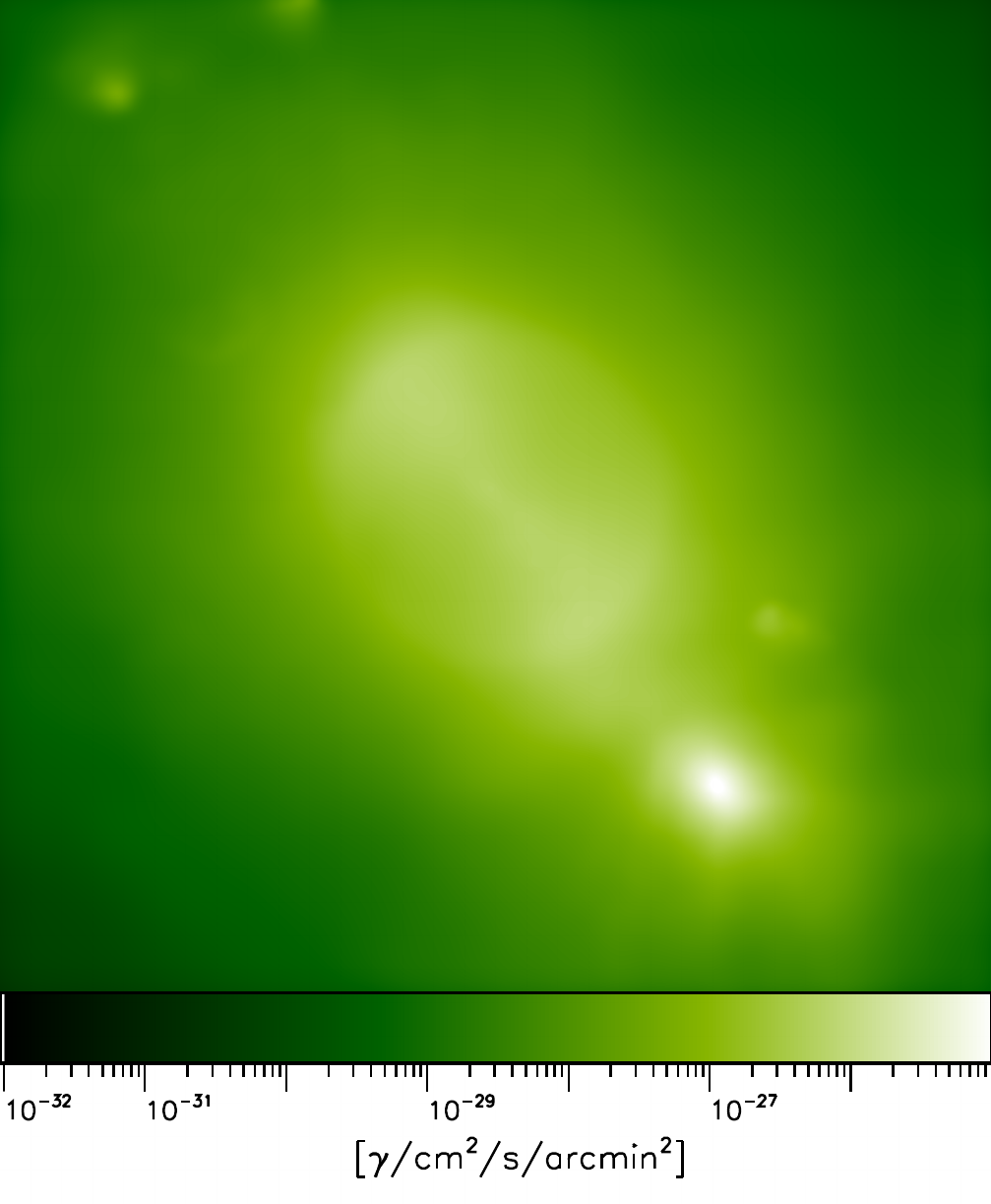}
\caption{Synthetic radio maps of the simulated Coma cluster (4 Mpc a side).
Upper panels are synchrotron maps at $1.4\,\mathrm{GHz}$, lower panels are 
$\gamma$-ray maps at $E_{\gamma} > 100\,\mathrm{GeV}$. The four models displayed 
from left to right are:
constant CRp energy fraction ({\it Model 1}), scaled CRp energy fraction ({\it Model 2}), 
scaled magnetic field ({\it Model 3}) and fitted CRp energy fraction ({\it Model 4}).
The overall normalization of the CRp energy fraction is chosen
to match the observed total radio luminosity 
of Coma. }\label{mapscoma}
\end{figure*}

In this letter we focus on hadronic models that can be implemented
in present simulations.
The spectrum of CRp was taken to be a power law, 
$N(E)=K_p E^{-\alpha_p}$, and the spectrum of the secondary
electrons resulting from p-p collisions was calculated under
stationary conditions considering synchrotron and inverse Compton losses
\citep[][and ref. therein]{2005MNRAS.363.1173B}.
Following \citet{2000A&A...362..151D} we adopted 
the slope of the spectral index of the Coma halo, $\alpha=1.25$,
obtain with $\alpha_p=2.6$ for the hadronic models 
(this also accounts for the logarithmic 
increase of the p-p cross-section with proton energy).
Such CRp collisions also produce $\gamma$-rays from the
decay of secondary neutral-pions, which are a direct measure of the 
CRp and of the unavoidable secondary electron injection process
into the ICM.
We calculate the $\gamma$-ray flux from our simulated Coma cluster
by adopting $\alpha_p=2.6$ and the formalism described in
\citet{2004A&A...413...17P}.

The spatial distribution of CRp in our MHD simulations is a free
parameter, so we follow four approaches chosen to encompass a range
suggested by theoretical and observational findings, but restricting
to large magnetic fields in the ICM that
provide the most favourable case for hadronic models.
In {\it Model 1} we assume that 
the energy density of the CR protons is 
a constant fraction of the thermal energy density of the ICM. 
In {\it Model 2} we adopt a radius-dependent ratio between CRp
and thermal energy density following results from cosmological
simulations of CRp acceleration at structure formation shocks by
\citet{2007MNRAS.378..385P}.
In both these models, we use 
the magnetic field strength and spatial distribution 
from our MHD simulations of the Coma cluster.
In {\it Model 3} we adopt the radius-dependent energy density of
CRp as in {\it Model 2} and {\it an artificial} magnetic field 
by assuming a radial scaling of the field strength in the form 
$B \propto \sqrt{\rho}$ within the radio emitting region (Figure 1).
In the last model, {\it Model 4}, we used the magnetic field from our 
MHD simulations, but leave the radial profile of the CRp energy density
in the simulated cluster free to vary to match the synchrotron brightness
profile measured for the Coma radio halo; in this case we limit the
CRp energy density  so that is stays smaller than the thermal energy density
of the ICM.

For all the models the overall normalization of the 
CRp energy density is choosen to match the observed radio luminosity
of the Coma radio halo at 1.4 GHz, $7.76 \times 10^{23}\,\mathrm{W/Hz}$ 
\citep{1997A&A...321...55D}. 

\section{The radio halo of Coma}

The Coma cluster hosts the prototype of
giant radio halos (eg. Giovannini et al.~1993), and in this section we compare
the radio properties from the simulated Coma cluster with the observed ones.
We calculate simulated radio images by using a map--making 
algorithm \citep{dolag2005}, that allows us to 
project the predicted emission of every SPH particle along the 
line of sight, 
considering an integration depth of $\pm$4Mpc around the center of the 
simulated Coma cluster. 
The resulting images of radio emission for our models are
shown in the upper panels of Figure \ref{mapscoma}.

\subsection{Radial profile and cosmic ray energy budget}

The radial distributions of the fraction between the energy density of CRp
and that of the 
thermal gas for {\it Models 1--4} are reported in the central panel
of Figure \ref{coma_profile}. 
Provided that the magnetic field strength in
the cluster-central regions is sufficiently large (i.e. $>$ 5 $\mu$G,
as in our cases), {\it Models 1--3}  
generate enough synchrotron luminosity at 1.4 GHz to match that of the 
Coma halo with reasonable requests in terms of energy density in the
CRp (about 1 to 10 \% of that of the thermal gas). 
However, a drawback of hadronic models discussed in the literature is 
that the radial distribution of the synchrotron emission 
generated from secondary electrons is expected to be
much steeper than that observed in radio halos, with most of
the radio luminosity generated in the cluster-core region 
\citep[e.g.][]{2004JKAS...37..493B}.

The right panel of Figure \ref{coma_profile} shows a comparison 
of the simulated and observed radial profile of the radio emission 
from the Coma cluster. 
In line with previous simulations \citep{2000A&A...362..151D} and
analytical expectations \citep[e.g.][]{2004JKAS...37..493B} we find that
assuming CRp contain a constant fraction of the energy
density of the thermal pool ({\it Model 1}) a profile  is produced that is
far too steep. 
Flatter radial profiles of the radio emission, and thus fairly 
extended radio halos, are generated only by {\it Models 2} and 3, 
yet the expected radio brightness at 0.5-1 Mpc distance from the
cluster-center is still $>$10 times fainter than that observed
in the Coma halo.
This picture also arises from the point-to-point scatter plot between
radio surface brightness and the X--ray brightness from
thermal emission (left panel in figure \ref{coma_spectrum}); the solid line 
is the observed correlation obtained by \citet{2001A&A...369..441G}.

\noindent
To reproduce the radial profile as measured at 1.4 GHz by \citet{1997A&A...321...55D} 
out to 0.5-1 Mpc distance, in {\it Model 4} 
we allow the energy density of CRp to vary with distance from the 
cluster center.
The resulting radial distribution of the 
amount of energy in CRp relative to that of the thermal pool is shown 
in the central panel of Figure \ref{coma_profile} and highlights an
expected energetic problem of hadronic models: to considerably
increase the synchrotron emissivity at large distance from cluster 
center, where the generation of secondary particles is inefficient 
due to the small number density of (target) thermal protons, a secondary model requires that 
the energy  density of CRp be extremely large, comparable with 
(or even larger than) that of the thermal pool.
Considering the profile measured with high sensitivity by Westerbork observations
at 330 MHz \citep[green, dashed line in the right panel of figure 
\ref{coma_profile},][]{2001A&A...369..441G} makes this point even more problematic.

\begin{table}
\centering
\begin{tabular}{c|c|c|c} 
\hline
Mission		& Veritas (diffuse) & ($r<1.3$ Mpc) & Egret \\
Model           & ($>0.1$TeV)	    & ($>0.1$TeV)	  & ($>0.1$GeV)  \\
\hline
observed        & $<2.0 \times 10^{-12}$  &  ---- & $<3.8 \times 10^{-8}$ \\
\hline
{\it Model1}	& $6.3 \times 10^{-14}$	& $6.6 \times 10^{-14}$	& $1.1 \times 10^{-9}$	\\
{\it Model2}	& $6.3 \times 10^{-14}$	& $7.8 \times 10^{-14}$	& $1.2 \times 10^{-9}$	\\
{\it Model3}	& $6.4 \times 10^{-14}$	& $8.0 \times 10^{-14}$	& $1.3 \times 10^{-9}$	\\
{\it Model4}	& $1.6 \times 10^{-13}$	& $3.7 \times 10^{-13}$	& $6.0 \times 10^{-9}$	\\
\hline
\end{tabular}
\caption{Expected limits of $\gamma$-ray
  emission from the simulated Coma cluster in
  $\gamma/\mathrm{s}/\mathrm{cm}^{2}$, excluding the substructure on the lower right. For the three models of CR
  distribution we show the limits in energy ranges,
  corresponding to recent experiments. } \label{table_limits}
\end{table}

\subsection{The Spectrum}\label{szcoma}

The radio spectrum of the Coma halo shows a steepening at higher
frequencies \citep{2003A&A...397...53T} that has been interpreted as a signature
of stochastic reacceleration of the emitting electrons 
\citep{1987A&A...182...21S,2001MNRAS.320..365B}.
On the other hand, 
it has also been argued that the steepening is due to incorrect 
estimation of the flux of the radio halo at the highest frequencies 
due to the Sunyaev-Zeldovich (SZ) decrement 
\citep[e.g.][]{2002A&A...396L..17E}.
The latter results in a flux reduction of the cosmic microwave background
photons in the cluster region by Compton scattering with the 
cluster-thermal electrons, which diminishes the radio flux at higher 
frequencies, although other authors conclude that 
this is not sufficient to explain the spectral steepening  
in the Coma halo \citep{2004A&A...424..773R,2004JKAS...37..493B}.

Our numerical simulations can be used to investigate 
SZ--decrement effect on the synchrotron spectral properties.
Figure \ref{coma_spectrum} (central panel) shows the radio and inverse 
Compton spectrum from the SZ effect inside the cluster gas. 
The red line is the prediction of the total flux from our models, 
where the SZ signal is extracted from the region of the radio
halo at 1.4 GHz, $\approx$500 kpc in radius.
The deviation from a pure power law at higher frequencies
is due to the SZ flux decrement (the dotted line marks the corresponding
power law with $\nu^{1.25}$), while the flattening at lower frequencies is due to the
energy-dependent cross-section of p-p collisions.
We plot the absolute of the IC flux, although 
it corresponds to negative flux at 
frequencies smaller than $\approx 2\times 10^{5}\,\mathrm{MHz}$. 
In black we show observations 
from \citet{2003A&A...397...53T} and \citet{2002ApJ...580L.101B} of
the Coma cluster in the radio and microwave band, respectively. 
We also evaluated the fluxes integrated over the relevant sizes according to the individual
observations (red symbols). 
The observed shape of the radio spectrum, 
with the steepening at $2\times 10^{3}\,\mathrm{MHz}$, 
cannot be explained by our secondary models, even including the 
SZ decrement for these frequencies, although the simulated
SZ signal almost perfectly fits observations in the microwave regime. 
The blue dashed line shows the inverse Compton decrement, 
assuming an isothermal beta
model for the thermal gas of Coma with emission region of $5$ Mpc radius
(as done in \citet{2002A&A...396L..17E}),
which would be needed to cause the steepening at 5 GHz. 
This size is about one 
order of magnitude more extended than the radio emission.
Additionally we show in green the expected spectrum 
from a stochastic electron-acceleration model 
\citep{1987A&A...182...21S}.

\section{$\gamma$-ray spectrum and limits}

\begin{figure*}
\centering
\includegraphics[width=0.33\textwidth]{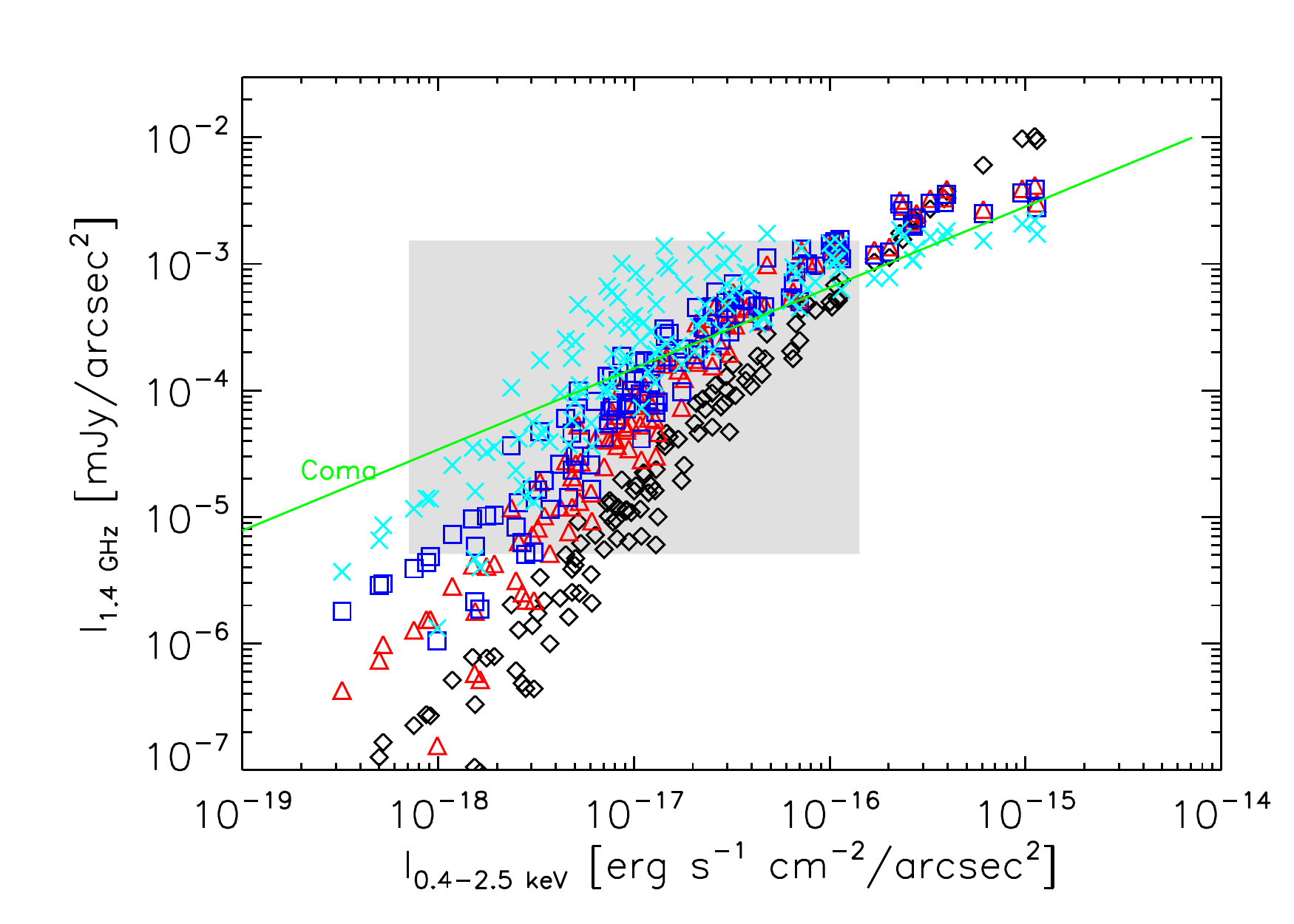}
\includegraphics[width=0.33\textwidth]{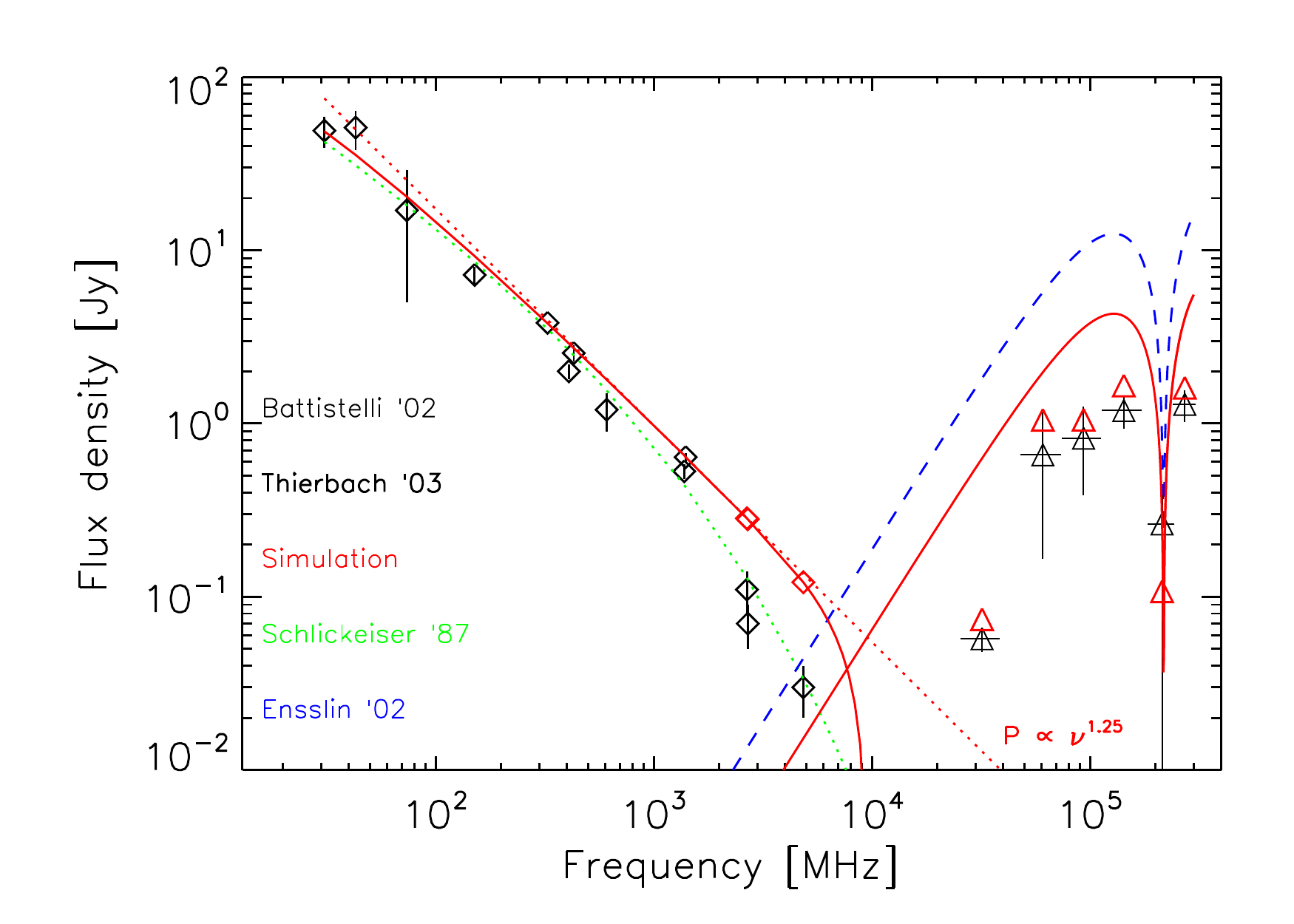}
\includegraphics[width=0.33\textwidth]{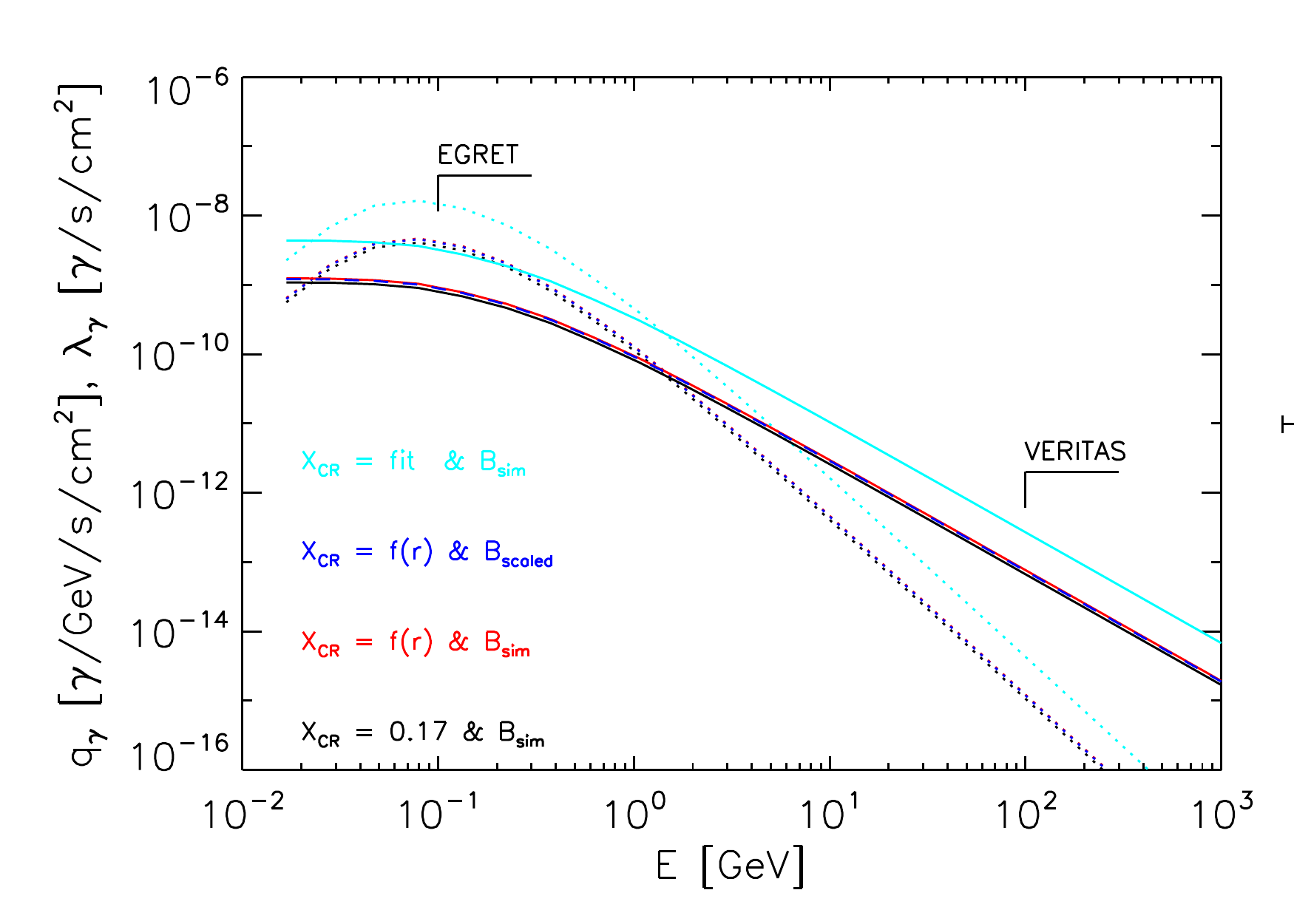}
\caption{The left panel shows the Radio Flux of the simulated Coma cluster at
$1.4\,\mathrm{GHz}$ versus X-ray surface brightness. We plot the
constant model in black, the model with scaled CRp in red, the one
with scaled MF and CRp in blue and the one with fitted CRp in magenta. 
The solid line is taken from 
the observations by \citet{2001A&A...369..441G}.
The middle panel shows the observed spectrum of the Coma cluster 
(synchrotron emission: black diamonds, \citet{2003A&A...397...53T};  
inverse Compton: black triangles, \citet{2002ApJ...580L.101B}).
Red lines are from the simulation and the red symbols are 
evaluated within the same aperture as the according observations.
See text for more details.
The right panel shows the expected $\gamma$-ray spectrum from the 
simulated Coma cluster for the our different models for the CRp 
distribution and the integrated
$\gamma$-ray spectrum (within $r<1.3$ Mpc) with the current upper limits
from EGRET and VERITAS 
\citep{2003ApJ...588..155R,2008AIPC.1085..569P}. 
}\label{coma_spectrum}
\end{figure*}

Maps of the predicted $\gamma$-ray emission of our simulated Coma cluster are 
shown in the lower panels of Figure \ref{mapscoma}. In the right panel of 
Figure \ref{coma_spectrum} we show the differential and
integral $\gamma$-ray flux as function of energy of the photons.
Here we also included the observed limits from VERITAS and EGRET 
\citep{2003ApJ...588..155R,2008AIPC.1085..569P}, see Table \ref{table_limits}.

We find that the $\gamma$-ray emission produced in the case
of {\it Model 4}, to allow a reasonable match with the
Deiss et al. profile of the Coma halo (at least at $<$ 0.7-1 Mpc distance
from cluster center), is a factor $\sim$6 below present upper limits.
As well, matching the radial profile of the Coma halo
as measured by high sensitivity observations at 330 MHz would require
an even larger energy budget of CRp with respect to that in 
{\it Model 4} and consequently a larger $\gamma$-ray emission.
Thus as soon as future $\gamma$-ray observations will reach sensitivities
slighly deeper than present upper limits they will allow complementary
tests to the hadronic origin of the radio emitting electrons 
in the Coma cluster. For instance, the expected sensitivity of the
Fermi Gamma Ray Telescope  will be sufficient to
constrain {\it Model 4}.

\section{Conclusions}\label{conclusions}

Based on a constrained, cosmological MHD simulation of the local universe 
we investigated the predicted properties of the radio halo of the Coma cluster 
within the framework of the hadronic model. 
We follow four approaches, chosen to span the reasonable range suggested 
by theoretical and observational findings, and focusing on the
case of high magnetic field values that represents the most favorable
way for hadronic models. 

\noindent
Our main conclusions are:
\begin{itemize}
\item In agreement with previous findings, hadronic models may
produce the synchrotron radio luminosity of the Coma halo with
energy densities of CRp between 1--10\% of the thermal ICM.
However the radial brightness profile of the generated synchrotron
emission is much steeper than seen in observations of the Coma halo,
consequently the simulated radio halos come out much smaller than the 
observed one. This also leads to a slope in the thermal X--ray vs.
radio brightness point-to-point correlation that is significantly
steeper than the observed one.
\item The observed flat radial brightness profile and the fairly large
extent of the observed radio halo can only be obtained with hadronic
models by strongly increasing the CRp energy density outside the core of the
cluster. In this case, however, the resulting CRp energy density at
$\approx$ 1Mpc (e.g. the rim of the observed radio halo) would
equal (or exceed) that of the thermal ICM.
\item Our simulated Coma cluster matches almost perfectly the observed SZ decrement.
But this SZ decrement is still not enough to explain the spectral steepening 
observed in the spectrum of the radio halo at large frequencies; contrary to previous claims in literature.
\end{itemize}

\noindent
In summary, we find that the hadronic model is disfavored 
by current observations of the Coma cluster.
In principle the energy problem
might be alleviated by if the magnetic field strength 
were almost constant with cluster radius up to the rim of the observed
radio halo \citep[see][]{2004A&A...413...17P}, however such models
are not among the best-fitting ones inferred from present RM
observations.
At the same time we have shown that hadronic models cannot explain
the spectral steepening of the Coma radio halo: regardless of the model 
assumptions, an SZ effect matching observations produces a negligible
SZ decrement at 2.7 and 5 GHz in the region of the halo.

The $\gamma$-ray flux generated by the hadronic interaction in case
of {\it Model 4}, which alone matches the observed radial 
profile is only a factor $\sim$6 below the current limits.
Considering that in case of lower magnetic fields (and/or assuming
the more extended profile of the Coma halo from high sensitivity 
330 MHz observations) the expected $\gamma$-ray flux should significantly 
increase, we conclude that
incoming $\gamma$-ray observations will shortly provide 
additional constraints to the models.

\section{Acknowledgements}
J.Donnert kindly acknowledges the support of ESF/Astrosim Exchange grant 
2065 and thanks the INAF/IRA in Bologna for its hospitality. GB, RC and AB 
acknowledge partial support from PRIN-INAF2007 and ASI-INAF I/088/06/0.

\bibliographystyle{mn2e} 
\bibliography{master}

\label{lastpage}
\end{document}